\documentclass[twocolumn]{autart}    

\usepackage{amsmath,amssymb,amsfonts,bm}
\usepackage{graphicx}
\usepackage{algorithm}
\usepackage{algorithmicx,algpseudocode}
\usepackage{cite}
\usepackage{color}
\let\theoremstyle\relax
\usepackage{amsthm}
\theoremstyle{plain}
\newtheorem{theorem}{Theorem}[section]
\newtheorem{lemma}[theorem]{Lemma}

\theoremstyle{definition}
\newtheorem{definition}[theorem]{Definition}
\theoremstyle{remark}
\newtheorem*{remark}{Remark}
\usepackage{wrapfig}

\begin{document}

\begin{frontmatter}

\title{Non-Gaussian Bayesian Filtering by Density Parametrization Using Power Moments} 

\author[Guangyu]{Guangyu Wu}\ead{chinarustin@sjtu.edu.cn},    
\author[Anders]{Anders Lindquist}\ead{alq@kth.se}              

\address[Guangyu]{Department of Automation, Shanghai Jiao Tong University, Shanghai, China}  
\address[Anders]{Department of Automation and School of Mathematical Sciences, Shanghai Jiao Tong University, Shanghai, China}


\begin{abstract}                          
    Non-Gaussian Bayesian filtering is a core problem in stochastic filtering. The difficulty of the problem lies in parameterizing the state estimates. However the existing methods are not able to treat it well. We propose to use power moments to obtain a parameterization. Unlike the existing parametric estimation methods, our proposed algorithm does not require prior knowledge about the state to be estimated, e.g. the number of modes and the feasible classes of function. Moreover, the proposed algorithm is not required to store massive parameters during filtering as the existing nonparametric Bayesian filters, e.g. the particle filter. The parameters of the proposed parametrization can also be determined by a convex optimization scheme with moments constraints, to which the solution is proved to exist and be unique. A necessary and sufficient condition for all the power moments of the density estimate to exist and be finite is provided. The errors of power moments are analyzed for the density estimate being either light-tailed or heavy-tailed. Error upper bounds of the density estimate for the one-step prediction are proposed. Simulation results on different types of density functions of the state are given, including the heavy-tailed densities, to validate the proposed algorithm. 
\end{abstract}

\end{frontmatter}

\section{Introduction}
\label{Introduction}
Stochastic filtering theory has been a fundamental topic in several areas including controls and signal processing for years and is applied in the various engineering and scientific areas, including communications, machine learning, neuroscience, economics, finance, political science, and many others. Pioneered by Norbert Wiener \cite{wiener1964extrapolation} and Andrey N. Kolmogorov \cite{kolmogorov1978stationary, shiryayev1992interpolation} in the 1940s, and promoted by Bode and Shannon \cite{bode1950simplified}, Zadeh and Ragazzini \cite{1950An} and others, a milestone of stochastic filtering theory was achieved by Rudolf E. Kalman \cite{kalman1961new, re1960new} in the 1960s. The Kalman filter (KF) consists of an iterative measurement-time update process. In the time-update step, the one-step ahead prediction of state is calculated; in the measurement-update step, the correction to the estimate of state according to the current observation is calculated. Moreover, the Kalman filter is indeed a time-variant Wiener filter \cite{anderson1971kalman}.

Even though the Kalman filter was originally derived with the orthogonal projection method under the LQG procedures, it has a decent Bayesian interpretation and can be derived within a Bayesian framework. Indeed, the Kalman filter can be regarded as a class of Bayesian filters of which the probability density function of the states and noises are Gaussian. Without exaggeration, the research on Bayesian filtering is inspired by the development of Kalman filtering. One of the first explorations of iterative Bayesian estimation is found in Ho and Lee’s paper \cite{ho1964bayesian}, where the principle and procedure of Bayesian filtering are specified. Sprangins \cite{spragins1965note} discussed the iterative application of Bayes rule to sequential parameter estimation. Lin and Yau \cite{lin1967bayesian} and Chien and Fu \cite{chien1967bayesian} discussed Bayesian approach to optimization of adaptive systems.

In practical situations, the state of the system and the noises do not always follow the Gaussian distribution. In the filtering problem in econometrics, for example in the analysis of financial time series, the distributions of the noises have heavy tails, where the Gaussian distribution does not apply. Moreover, when the probability density of the state is multi-modal, it is not feasible to estimate it with a Gaussian distribution. In the scenarios where the Kalman filter does not apply, we consider the Bayesian filter naturally due to its relaxation on the type of the densities.

However, there is always a tradeoff in stochastic filtering. For the Kalman filter, the densities being Gaussian makes it feasible to obtain an analytic form of solution of the convolution in the time update step. Bayesian filter does not require the densities to be Gaussian, but the solvability of the convolution (integration) is not guaranteed. In previous research results, several numerical methods have been proposed to obtain an analytic solution to the integration in the time update step, including Gaussian/Laplace approximation \cite{mackay1998choice}, iterative quadrature \cite{freitas1999bayesian, wang1978optimal, kushner2000nonlinear}, Gaussian sum approximation \cite{sorenson1971recursive, alspach1972nonlinear} and state-space calculus \cite{hanzon2001state}. These are the parametric methods for parameterizing the probability density function in a continuous form. However, the flexibility of those methods is too limited, which makes it difficult to apply the methods to a wide class of density functions. Moreover, quantitative approximation performance analyses of the methods, e.g. an error upper bound of estimation, have not been proposed yet, which severely decreases the value of these methods in practical use. 

Meanwhile, there are also several methods which characterize the density in a discrete form of the function, including mulitgrid method and point-mass approximation \cite{cai1995adaptive, bucy1971digital} and Monte Carlo sampling approximation \cite{handschin1969monte, christian1999monte}. These nonparametric methods impose no prior constraints on the density functions which seem to enjoy the maximum flexibility, however the tradeoff is also very severe: quite a number of probability values at discrete states need to be stored, and the continuity of the density is sacrificed. It means that when given an arbitrary state, we are always not able to obtain its value of probability. At the same time, as to guarantee the computation efficiency of these algorithms in filtering, resampling is widely used in the filters to avoid depletion of particles with small probability values. In some applications, we only consider the states with significant values of probability; however the states of small values of probability are extremely important, e.g. in financial applications. In conclusion, the discrete methods for density characterization are intrinsically infeasible in tackling the problem where the states with less significant values of probability still have dominant impact on the filtering problem.

Let us return to the methods parameterizing the density function in a continuous form. The Kalman filter estimates the first two orders of moments, but it is natural to consider using the moments of higher orders for filtering. Unfortunately the classes of density functions, which the filters are able to treat, are still very limited in previous papers \cite{srinivasan1970state}.

In this paper, we first formulate the non-Gaussian Bayesian filtering problem in Section 2. We propose to use the higher order moments to characterize the density function. The density surrogate is also defined. A construction of the density surrogate, i.e., parametrization of the density function is proved to exist and proposed in Section 3. In doing this, we follow the procedure of  \cite{georgiou2003kullback}, where a large class of trigonometric moment problems are solved by minimizing a Kulback-Leibler type criterion. Indeed, this theory can be modified to the power moment problem, which is what is needed here. The parameters of the model can be determined by a convex optimization scheme and the map from the parameters to the power moments is proved to be a homeomorphism. It ensures that the gradient-based optimization algorithms can be applied to determining the parameters of the density surrogate. The parametrization is in terms of a prior density $\theta$, and in Section 4, we give a sufficient and necessary condition on $\theta$ for the density surrogate $\hat{\rho}$ to have all power moments to exist and be finite. With the prior distribution selected as a sub-Gaussian distribution, the estimated moments of the density surrogate are proved to be asymptotically unbiased from the true ones when using the density surrogate with the highest order of the moments used tending to infinity. By selecting a sufficient large $n$, we have that the estimated moments are approximately the true ones, i.e., using the density surrogate will not bring significant cumulative errors to the moment estimation of the subsequent filtering steps. To our knowledge, the asymptotic unbiasedness of the statistics of the estimated density function has not previously been proved for the Bayesian filters of which the state is a continuous function, and cumulative errors are always hard to predict. Moreover, an upper bound of the approximation error in the sense of total variation distance is proposed, which has not previously been done for the Bayesian filters with no prior constraints on the classes of the densities. Error upper bounds of the density estimate for the probability of subsets of the real line are also proposed. Simulation results of different classes of density functions, including heavy-tailed ones, are given in Section 5 to validate the performance of our filter.

\section{Problem formulation}
\label{ProblemFormulation}
In this paper, following \cite{hanzon2001state}, we consider the non-Gaussian filtering problem for the first order system
\begin{equation}
\label{System}
    \begin{aligned}
    x_{t+1} &=f_{t} x_{t}+\eta_{t} \\
    y_{t} &=h_{t} x_{t}+\epsilon_{t}
\end{aligned}    
\end{equation}
$t=0,1,2, \ldots$. The state $x_{t}$ is a random variable defined on $\mathbb{R}$, and $f_{t}, h_{t}$ are assumed to be known real numbers. The system noise $\eta_{t}$ is a random variable defined on $\mathbb{R}$, which can be either continuous or discrete. When $\eta_{t}$ is continuous, the probability density function is assumed to be non-Gaussian. The observation noise $\epsilon_{t}$ is assumed to be a Lebesgue integrable function. The noises are assumed to be independent from each other, and their densities are denoted as $\rho_{\eta_{t}}$ and $\rho_{\epsilon_{t}}$. 

We adopt the Bayesian filter as used in \cite{hanzon2001state}. Denoting the collection of observations $y_{t}, y_{t-1}, \cdots, y_{0}$ by $\mathcal{Y}_{t}$, 
the conditional densities of the measurement and time updates are given by the following

\noindent \textbf{Measurement update}: For $t=0$,
\begin{equation}
\begin{aligned}
    \rho_{x_{0} \mid \mathcal{Y}_{0}}(x)& =\frac{\rho_{y_{0} \mid x_{0}}\left(y_{0}\right) \rho_{x_{0}}(x)}{\int_{\mathbb{R}}\rho_{y_{0} \mid x_{0}}\left(y_{0}\right) \rho_{x_{0}}(x)dx} \\
    & =\frac{\rho_{\epsilon_{0}}\left(y_{0}-h_{0} x\right) \rho_{x_{0}}(x)}{\int_{\mathbb{R}}\rho_{\epsilon_{0}}\left(y_{0}-h_{0} x\right) \rho_{x_{0}}(x)dx};
\label{Update1}
\end{aligned}
\end{equation}
for $t \geq 1$,
\begin{equation}
\begin{aligned}
\rho_{x_{t} \mid \mathcal{Y}_{t}}(x) & =\frac{\rho_{y_{t} \mid x_{t}}\left(y_{t}\right) \rho_{x_{t} \mid \mathcal{Y}_{t-1}}(x)}{\int_{\mathbb{R}}\rho_{y_{t} \mid x_{t}}\left(y_{t}\right) \rho_{x_{t} \mid \mathcal{Y}_{t-1}}(x)dx}\\
& =\frac{\rho_{\epsilon_{t}}\left(y_{t}-h_{t} x\right) \rho_{x_{t} \mid \mathcal{Y}_{t-1}}(x)}{\int_{\mathbb{R}}\rho_{\epsilon_{t}}\left(y_{t}-h_{t} x\right) \rho_{x_{t} \mid \mathcal{Y}_{t-1}}(x)dx}, x \in \mathbb{R}.
\label{Update2}
\end{aligned}
\end{equation}

\noindent \textbf{Time update}: For $t \geq 0$,
\begin{equation}
\begin{aligned}
    \rho_{x_{t+1} \mid \mathcal{Y}_{t}}(x) & =\left(\rho_{f_{t} x_{t} \mid \mathcal{Y}_{t}} * \rho_{\eta_{t}}\right)(x)\\
    & =\int_{\mathbb{R}} \rho_{x_{t} \mid \mathcal{Y}_{t}}\left(\frac{\xi}{f_{t}}\right) \rho_{\eta_{t}}(x-\xi) d\xi.
\label{Prediction}
\end{aligned}
\end{equation}

As are derived in \eqref{Update1}, \eqref{Update2} and \eqref{Prediction}, $\rho_{x_{t} \mid \mathcal{Y}_{t}}(x)$ and $\rho_{x_{t+1} \mid \mathcal{Y}_{t}}(x)$ are evaluated at $x$. In the following part of this paper, we write $\rho_{x_{t} \mid \mathcal{Y}_{t}}, \rho_{x_{t+1} \mid \mathcal{Y}_{t}}$ for simplicity, if there is no ambiguity. Even though the densities are all non-Gaussian, the measurement update (\ref{Update2}) is a multiplication of two densities, therefore can be performed easily. But it is not always possible to obtain an explicit form of the one-step prediction in (\ref{Prediction}) when the densities are not Gaussian \cite{chen2003bayesian}. Now the problem becomes the approximation of $\rho_{x_{t+1} \mid \mathcal{Y}_{t}}$. However, we notice that the power moments of $\rho_{x_{t+1} \mid \mathcal{Y}_{t}}$: i.e.,  
\begin{equation}
\label{sigma}
 \sigma_{k, t+1} := \int_{\mathbb{R}} x^k\rho_{x_{t+1}\mid\mathcal{Y}_t}dx =\mathbb{E}\left ( x_{t+1}^{k}|\mathcal{Y}_{t} \right ),
\end{equation}
are easy to obtain. In fact, by \eqref{System},
\begin{equation*}
\begin{aligned}
    \sigma_{k, t+1}
    = & \mathbb{E}\left ( \left( f_{t}x_{t}+\eta_{t}\right)^{k}|\mathcal{Y}_{t} \right ) \\
    = & \mathbb{E}\left ( \sum_{j=0}^{k}\left ( \begin{matrix} k \\j\end{matrix} \right )
     f^{j}_{t}x^{j}_{t}\cdot\eta^{k-j}_{t}|\mathcal{Y}_{t} \right ) \\
= & \sum_{j=0}^{k}\left ( \begin{matrix} k \\j \end{matrix} \right ) 
\mathbb{E}\left(f^{j}_{t}x^{j}_{t}\cdot\eta^{k-j}_{t}|\mathcal{Y}_{t} \right ).
\end{aligned}
\end{equation*}
 Therefore, since $x_{t}$ and $\eta_{t}$ are independent,
\begin{equation}
\label{MomentUpdate}
\sigma_{k, t+1}=\sum_{j = 0}^{k}\left ( \begin{matrix} k\\j\end{matrix} \right ) f_{t}^{j}\mathbb{E}\left ( x_{t}^{j} |\mathcal{Y}_{t} \right )\mathbb{E}\left ( \eta_{t}^{k-j} \right )
\end{equation}
for $k = 1, \cdots, 2n$.
Inspired by the method of moments, we propose to use the truncated power moments to estimate $\rho_{x_{t+1} \mid \mathcal{Y}_{t}}$.

Previous research has been focusing on density approximation. In the Kalman filter (and its extended forms, e.g. extended Kamlan filter and unscented Kalman filter), approximation of the one-step prediction is a parametric estimation problem, which is done by estimating the first and second order moments. On the other hand, the particle filter is proposed as a non-parametric algorithm to tackle this problem. However, there is no convenient way of error analysis and its performance suffers from sample depletion.

In this paper, we propose a filter which not only admits treating the non-Gaussian state estimation problem but also provides an analytic error analysis to measure the performance of filtering. To this end, we define the probabilities with the identical truncated power moment sequence.

\medskip

\begin{definition}
A probability density function, whose first $2n$ power moments coincide with those of  the probability density $\rho$, is called an order-$2n$ density surrogate of $\rho$ and is denoted by $\rho^{2n}$.
\end{definition}
Denoting the density prediction as $\hat{\rho}$, we propose to perform each iteration of Bayesian filtering  with the density surrogate as in Algorithm \ref{Algo1}. 

\begin{algorithm}[htb] 
\caption{Bayesian filtering with density surrogate at time $t$.} 
\begin{algorithmic}[1]
\Require System parameters : $f_{t}, h_{t}$;
non-Gaussian densities : $\eta_{t}, \epsilon_{t}$;
prediction at time $t-1$ : $\rho_{x_{0}}(x) \text{ or } \hat{\rho}_{x_{t} \mid \mathcal{Y}_{t-1}}(x)$;
\Ensure
Prediction at time $t$ : $\hat{\rho}_{x_{t+1} \mid \mathcal{Y}_{t}}(x)$.
\State Calculate $\hat{\rho}_{x_{t}|\mathcal{Y}_{t}}$ by (\ref{Update1}) or (\ref{Update2}); 
\State Calculate $\sigma_{t}$ by (\ref{MomentUpdate});
\State Determine an order-$2n$ density surrogate $\rho^{2n}_{x_{t+1} \mid \mathcal{Y}_{t}}$, of which the truncated moment sequence is $\sigma_{t}$. The density estimate for the one-step prediction at time $t+1$ is then chosen as the density surrogate, i.e., $\hat\rho_{x_{t+1} \mid \mathcal{Y}_{t}} = \rho^{2n}_{x_{t+1} \mid \mathcal{Y}_{t}}$.
\label{AlgoSurrogate}
\end{algorithmic}
\label{Algo1}
\end{algorithm}

Now the problem comes down to constructing an order-$2n$ density surrogate. Since the domain of $\rho$ is $\mathbb{R}$, the problem becomes a Hamburger moment problem \cite{schmudgen2017moment}. In the next section, we will give a formal definition to the Hamburger moment problem we will treat and give a solution to the problem, i.e. a parametrization of the density surrogate.

\section{Parametrization of the density surrogate using power moments}
\label{parametrizationOf}

As seen from \eqref{sigma}, determining a  density $\hat{\rho}_{x_{t+1}\mid\mathcal{Y}_t}$ given the power moments is a (truncated) Hamburger moment problem,
For simplicity we omit the subscript $t$ in all terms.

\medskip

\begin{definition}
A sequence 
\begin{equation}
    \bar{\sigma}_{2n} = (\sigma_0, \sigma_1, \dots, \sigma_{2n})
\end{equation}
is a feasible $2n$-sequence, if there is a random variable $X$ with a probability density $\rho(x)$ defined on $\mathbb{R}$, whose moments are given by \eqref{MomentUpdate}, that is, 
\begin{equation}
    \sigma_{k} = \mathbb{E}\{X^{k} \} =\int_\mathbb{R}x^k\rho(x)dx, \quad k=0,1,\dots, 2n.
\label{MomentProblem}
\end{equation}
\label{definition31}
We say that any such random variable $X$ has a $\bar{\sigma}_{2n}$-feasible distribution. As to ensure the existence of $\rho$, the Hankel matrix
\begin{equation}
    \Sigma:=\left[\begin{array}{cccc}
    \sigma_{0} & \sigma_{1} & \ldots & \sigma_{n} \\
    \sigma_{1} & \sigma_{2} & \ldots & \sigma_{n+1} \\
    \vdots & \vdots & \ddots & \vdots \\
    \sigma_{n} & \sigma_{n+1} & \ldots & \sigma_{2n}
    \end{array}\right]
\label{HankelMat}
\end{equation}
needs to be positive definite, i.e., the sequence $\bar{\sigma}_{2n}$ needs to be a positive one.
\end{definition}

For Bayesian filtering, we need to propagate the density estimates throughout the filtering process, which makes it necessary to derive an analytic form of $\rho(x)$ with finitely many parameters. It is called the parametrization of the density function, which is essentially a dimension reduction problem. We emphasize that the solution to the problem is not unique and that there are in general infinitely many solutions. Next we proceed to describe these.

Observe that the moment conditions
\begin{equation}
\label{momentcondition}
\sigma_{k} =\int_\mathbb{R}x^k\rho(x)dx, \quad k=0,1,\dots,2n
\end{equation}
can be written in the matrix form
\begin{equation}
   \int_\mathbb{R} G(x)\rho(x) G^{T}(x)dx=\Sigma ,
\label{IntegG}
\end{equation}
where 
$$
G(x)= \begin{bmatrix}
1 & x & \cdots & x^{n-1} & x^{n}
\end{bmatrix}^{T},
$$
and $\Sigma$ is the Hankel matrix of the form \eqref{HankelMat}, of which the entries $\sigma_{k}, k = 0, \cdots, 2n$ are calculated by \eqref{MomentUpdate}. Consequently, we have an order $2n$ moment estimation problem as defined in Definition \ref{definition31}. Indeed, it is an Hankel matrix representation of the Hamburger moment problem.

We note that $\Sigma$ is a Hankel matrix hence symmetric. By \eqref{IntegG} we have that, for each $a\in\mathbb{R}_{n+1}/\{0\}$, $a'\Sigma a=\int a(x)^2\rho(x)dx$, where $a(x)=a'G(x)$. Since $a(x)^2$ is positive except possibly in a set of measure zero, $a'\Sigma a >0$ for all $a$, and by Definition 4.1.11 in \cite{horn2012matrix}, $\Sigma$ is positive definite. By Corollary 9.2 in \cite{schmudgen2017moment}, the positive definiteness of $\Sigma$ also shows that there exists at least one solution to the corresponding truncated moment problem.

Let $\mathcal{P}$ be the space of probability density functions on the real line with support there, and let $\mathcal{P}_{2n}$ be the subset of all $\rho\in\mathcal{P}$ which have at least  $2n$ finite moments (in addition to $\sigma_0$, which of course is 1). We note that the class of $\rho\in\mathcal{P}$ satisfying \eqref{IntegG} is nonempty since $\Sigma$ is positive definite ($\Sigma \succ 0$). In fact, $\Sigma$ is in the range of the linear integral operator
\begin{equation}
    \Gamma: \rho \mapsto \Sigma=\int_\mathbb{R} G(x) \rho(x) G^{T}(x)dx ,
\label{Map}
\end{equation}
which is defined on the space $\mathcal{P}_{2n}$. Since  $\mathcal{P}_{2n}$ is convex, then so is $\operatorname{range}(\Gamma)=\Gamma\mathcal{P}_{2n}$.

Let $\theta$ be an arbitrary prior density in $\mathcal{P}$ and consider the Kullback-Leibler (KL) (pseudo) distance
\begin{equation}
\label{KL}
\mathbb{KL}(\theta\|\rho)=\int_\mathbb{R} \theta(x) \log \frac{\theta(x)}{\rho(x)} dx
\end{equation}
between $\theta$ and $\rho$. Although not symmetric in its arguments, the KL distance is jointly convex.  It is widely used in density estimation \cite{hall1987kullback, li1999mixture, vapnik1999nature}. 

In \cite{georgiou2003kullback}, the KL distance was used as a distance measure between spectral densities. In this section, following the line of thought of \cite{georgiou2003kullback}, we introduce a parametrization of $\rho\in\mathcal{P}_{2n}$ satisfying \eqref{momentcondition},  which is induced by the KL distance. The results are very similar to those in \cite{georgiou2003kullback}, but since here we are dealing with a power moment problem rather than a trigonometric moment problem as in \cite{georgiou2003kullback}, important details of the proofs are different, so we need to proceed with care.

\medskip

\begin{theorem} \label{mainthm}
Let $\Gamma$ be defined by (\ref{Map}), and let
\begin{equation*}
\label{Lplus}
\mathcal{L}_{+}:=\left\{\Lambda \in \operatorname{range}(\Gamma) \mid G\left(x\right)^{T} \Lambda G\left(x\right)>0, x  \in \mathbb{R} \right\}.
\end{equation*}
Given any $\theta \in \mathcal{P}$ and any $\Sigma \succ 0$, there is a unique $\rho \in \mathcal{P}_{2n}$ that minimizes \eqref{KL}
subject to $\Gamma(\rho)=\Sigma$, i.e., subject to \eqref{IntegG}, namely 
\begin{equation}
    \hat{\rho}=\frac{\theta}{G^{T} \hat{\Lambda} G},
\label{MiniForm}
\end{equation}
where $\hat{\Lambda}$ is the unique solution to the problem of minimizing
\begin{equation}
    \mathbb{J}_\theta(\Lambda):=\operatorname{tr}(\Lambda \Sigma)-\int_\mathbb{R} \theta(x) \log\left[ G(x)^{T} \Lambda G(x)\right] dx
\label{LossFunc}
\end{equation}
over all $\Lambda \in \mathcal{L}_{+}$. Here $\operatorname{tr}(M)$ denotes the trace of the matrix $M$.
\end{theorem}

\begin{proof}
First form the Lagrangian
$$
L(\rho, \Lambda)=\mathbb{KL}(\theta \| \rho)+\operatorname{tr}(\Lambda(\Gamma(\rho)-\Sigma)),
$$
where $\Lambda \in\operatorname{range}(\Gamma)$ is the matrix-valued Lagrange multiplier, and consider the problem of maximizing the dual functional
\begin{equation}
\label{infL}
\Lambda \mapsto \inf _{\rho \in \mathcal{P}_{2n}} L(\rho, \Lambda).
\end{equation}
Clearly $\rho\mapsto L(\rho, \Lambda)$ is strictly convex, so to be able to determine the right member of \eqref{infL}, we must find a $\rho\in\mathcal{P}_{2n}$, for which the directional derivative $\delta L(\rho, \Lambda ; \delta\rho) =0$ for all relevant $\delta\rho$. This will further restrict the choice of $\Lambda$.
Setting 
\begin{equation}
\label{q}
q(x):=G(x)^T\Lambda G(x),
\end{equation}
we have
$$
L(\rho, \Lambda)=\int_\mathbb{R} \theta(x)\log \frac{\theta(x)}{\rho(x)}dx +\int_\mathbb{R}q(x)\rho(x)dx-\operatorname{tr}(\Lambda \Sigma),
$$
with the directional derivative
$$
\delta L(\rho, \Lambda ; \delta\rho)=\int_\mathbb{R} \delta\rho(x)\left(q(x)-\frac{\theta(x)}{\rho(x)}\right)dx ,
$$
which has to be zero at a minimum for all variations $\delta\rho$. Clearly this can be achieved only if $q(x)=\theta(x)/\rho(x)$ for all $x\in\mathbb{R}$. To complete the proof of Theorem~\ref{mainthm}, we need some more preliminary results, but let us first make an important observation.

\medskip

\begin{remark}The parametrization of the density surrogate by the Hankel matrix restricts the highest order of the terms of the denominator to be even, i.e., $2n$. Indeed, it is the necessary condition for a polynomial to be always positive everywhere on $\mathbb{R}$. A polynomial for which the order of the highest order term is odd always has a real zero, and the value of the polynomial changes sign at that point. It makes constructing the density surrogate problematic.
\end{remark}

In particular, this requires the condition $q(x)>0$ for all $x\in\mathbb{R}$, so by \eqref{IntegG} and \eqref{q}, we obtain the constraint $\Lambda \in \mathcal{L}_+$, which is proved in the following lemma. 

\medskip

\begin{lemma}
$\Lambda \in \mathcal{L}_+$ only if $q(x)>0$.
\label{lemma33}
\end{lemma}

\begin{proof}
Since $\Lambda \in \mathcal{L}_+$, we write $\Lambda$ as
$$
   \int_\mathbb{R} G(y)\psi(y) G^{T}(y)dx=\Lambda,
$$
where $\psi \in \mathcal{P}_{2n}$. Therefore we have
$$
   G^{T}(x)\int_\mathbb{R} G(y)\psi(y) G^{T}(y)dx G(x)=G^{T}(x)\Lambda G(x) = q(x),
$$
and hence
$$
q(x)=\int_\mathbb{R} \varphi_x(y)\psi(y) dy,
$$
for each $x$, where
$$
\varphi_x(y)=\left[ G^T(x)G(y)\right]^2 \geq 0.
$$
However, for each fixed $x$,  $\varphi_x(y)$ is a polynomial such that $\varphi_x(0)=1$, and hence  $\varphi_x(y)=0$ at most in a finite number of $y$. Consequently, since $\psi(y)>0$, we have $q(x)>0$ for all $x$.
\end{proof}

Moreover, a possible minimizer must have the form
$$
\rho=\frac{\theta}{q},
$$
and the dual functional must be
\begin{displaymath}
L\left(\frac{\theta}{q},\Lambda\right)=-\mathbb{J}_{\theta}(\Lambda)+\int_\mathbb{R}\theta(x)dx,
\end{displaymath}
where $\mathbb{J}_{\theta}(\Lambda)$ is given by \eqref{LossFunc}. Therefore the dual problem amounts to minimizing $\mathbb{J}_{\theta}(\Lambda)$ over $\mathcal{L}_+$. To conclude the proof of Theorem~\ref{mainthm} we need the following theorem, which will be proved in the following part of this section.

\medskip

\begin{theorem}
The functional $\mathbb{J}_{\theta}(\Lambda)$ has a unique minimum $\hat{\Lambda} \in\mathcal{L}_{+}$. Moreover
$$
\Gamma\left(\frac{\theta}{G^{T} \hat{\Lambda} G}\right)=\Sigma.
$$
\label{theorem34}
\end{theorem}

By this theorem 
$$
\hat{\rho}=\frac{\theta}{\hat{q}}, \quad \text{where $\hat{q}=G^{T} \hat{\Lambda} G$},
$$
belongs to $\mathcal{P}_{2n}$ and is a stationary point of $\rho \mapsto L(\rho, \hat{\Lambda})$, which is strictly convex. Consequently
$$
L(\hat{\rho}, \hat{\Lambda}) \leq L(\rho, \hat{\Lambda}), \quad \text { for all } \rho \in \mathcal{P}_{2n}
$$
or, equivalently, since $\Gamma(\hat{\rho})=\Sigma$,
$$
\mathbb{KL}(\theta\| \hat{\rho}) \leq \mathbb{KL}(\theta \| \rho)
$$
for all $\rho\in \mathcal{P}_{2n}$ satisfying the constraint $\Gamma(\rho)=\Sigma$. The above holds with equality if and only if $\rho=\hat{\rho}$. 
This completes the proof of Theorem \ref{mainthm}.
\end{proof}

To prove Theorem \ref{theorem34}, we need to consider the dual problem to minimize $\mathbb{J}_{\theta}(\Lambda)$ over $\mathcal{L}_{+}$.

\medskip

\begin{lemma}
Any stationary point of $\mathbb{J}_{\theta}(\Lambda)$ must satisfy the equation
\begin{equation}
    \omega(\Lambda)=\Sigma ,
    \label{Omega}
\end{equation}
where the map $\omega: \mathcal{L}_{+} \mapsto \mathcal{S}_{+}$ between $\mathcal{L}_{+}$ and $\mathcal{S}_{+}:=\{\Sigma \in \operatorname{range}(\Gamma)\mid \Sigma \succ 0\}$ is defined as
$$
\omega: \; \Lambda \mapsto \int_\mathbb{R} G(x) \frac{\theta(x)}{q(x)} G(x)^{T}dx
\label{omega}
$$
with $q$ defined by \eqref{q}.
\end{lemma}

\begin{proof}
From \eqref{LossFunc} and \eqref{q} we have
\begin{displaymath}
\mathbb{J}_{\theta}(\Lambda)= \operatorname{tr}\{\Lambda\Sigma\} -\int_\mathbb{R}\theta(x)\log q(x) dx,
\end{displaymath}
and therefore, using the fact that 
\begin{displaymath}
\delta q(\Lambda;\delta\Lambda)=G^T\delta\Lambda G = \operatorname{tr}\{\delta\Lambda GG^T\},
\end{displaymath}
we have the directional derivative
\begin{equation*}
\delta \mathbb{J}_{\theta}(\Lambda ; \delta \Lambda)=\operatorname{tr}\left(\delta \Lambda\left[\Sigma-\int_\mathbb{R} G(x) \frac{\theta(x)}{q(x)} G(x)^{T}dx\right]\right),
\label{FirstOrder}
\end{equation*}
which is zero for all $\delta \Lambda \in\operatorname{range}(\Gamma)$ if and only if \eqref{Omega} holds. This completes the proof.
\end{proof}

To prove Theorem \ref{theorem34}, we also need to establish that the map $\omega: \mathcal{L}_{+} \rightarrow \mathcal{S}_{+}$ is injective, establishing uniqueness, and surjective, establishing existence. In this way we prove that \eqref{Omega} has a unique solution, and hence that there is a unique minimum of the dual functional $\mathbb{J}_{\theta}$. We start with injectivity.

\medskip

\begin{lemma}
Suppose $\Lambda \in \text{range}(\Gamma)$. Then the map 
\begin{equation}
\label{Lambda2G'LambdaG}
\Lambda \mapsto G^{T} \Lambda G
\end{equation}
is injective.
\label{lemma36}
\end{lemma}

\begin{proof}
Since $\Lambda \in \text{range}(\Gamma)$, 
$$
\label{xx}
   \Lambda = \int_{\mathbb{R}} G(y) \psi(y) G^{T}(y)dy
$$
for some $\psi \in \mathcal{P}$.
Suppose $G^{T}\Lambda G = 0$. Then we have $\int_{\mathbb{R}} G^{T}(x)\Lambda G(x)dx = 0$, and therefore
$$
\begin{aligned}
    & \int_{\mathbb{R}} G^{T}(x)\Lambda G(x)dx \\
    = & \operatorname{tr} \left( \int_\mathbb{R} G(x)^{T}\int_\mathbb{R} G(y) \psi(y) G(y)^{T}dy\, G(x) dx \right) \\
    = & \int_\mathbb{R} \int_\mathbb{R} [G(x)^TG(y)]^2\psi(y) dxdy =0.
\end{aligned}
$$
Consequently we have $[G(x)^TG(y)]^2\psi(y) = 0$, for all $x,y \in \mathbb{R}$, which clearly implies that $\psi=0$, and hence that $\Lambda = 0$.
Consequently the map \eqref{Lambda2G'LambdaG} is injective, as claimed.
\end{proof}

To prove that $\omega: \mathcal{L}_{+} \rightarrow \mathcal{S}_{+}$ is injective, suppose that $\omega(\Lambda_1)=\omega(\Lambda_2)$ for some $\Lambda_1$ and $\Lambda_2$ in $\mathcal{L}_+$. We need to show that $\Lambda_1=\Lambda_2$. To this end, note that
$$\omega(\Lambda_1)-\omega(\Lambda_2)=\int_\mathbb{R}GG^T\frac{\theta}{q_1q_2}(q_2-q_1)dx=0,$$
where $q_1=G^T\Lambda_1G$ and $q_2=G^T\Lambda_2G$. In view of Lemma \ref{lemma33}, this implies that $q_1=q_2$, so by Lemma \ref{lemma36} this implies that $\Lambda_1=\Lambda_2$, establishing that $\omega$ is injective.


Next, we shall prove that $\omega: \mathcal{L}_{+} \mapsto \mathcal{S}_{+}$ is also surjective. To this end, we first note that $\omega$ is continuous and that both sets $\mathcal{L}_{+}$ and $\mathcal{S}_{+}$ are nonempty, convex, and open subsets of the same Euclidean space, and hence diffeomorphic to this space. For the proof of surjectivity we shall use  Corollary 2.3 in \cite{byrnes2007interior}, by which the continuous map $\omega$ is surjective if and only if it is injective and proper, i.e., the inverse image $\omega^{-1}(K)$ is compact for any compact $K$ in $\mathcal{S}_{+}$. (For a more general statement, see Theorem 2.1 in \cite{byrnes2007interior}.) Consequently it just remains to prove that $\omega$ is proper. To this end, we first note that $\omega^{-1}(K)$ must be bounded, since, as if $\|\Lambda\|\to \infty$,  $\omega(\Lambda)$ would tend to zero, which lies outside $\mathcal{L}_{+}$. Now, consider a Cauchy sequence in $K$, which of course converges to a point in $K$. We need to prove that the inverse image of this sequence is compact. If it is empty or finite, compactness is automatic, so suppose it is infinite. Then, since $\omega^{-1}(K)$ is bounded, there must be a subsequence $(\lambda_k)$ in $\omega^{-1}(K)$ converging to a point $\lambda\in\mathcal{L}_{+}$. It remains to show that $\lambda\in\omega^{-1}(K)$, i.e., $(\lambda_k)$ does not converge to a boundary point, which here would be $q(x)=0$. However this does not happen since then $\operatorname{det} \omega(\Lambda) \rightarrow \infty$, contradicting boundedness of $\omega^{-1}(K)$. Hence $\omega$ is proper. 

Therefore, the map $\omega: \mathcal{L}_{+} \rightarrow \mathcal{S}_{+}$ is proved to be homeomorphic, which completes the proof of Theorem \ref{theorem34}.

Consequently, the dual problem provides us with an approach to compute the unique $\hat{\rho}$ that minimizes the Kullback-Leibler distance $\mathbb{KL}(\theta \| \rho)$ subject to the constraint $\Gamma(\rho)=\Sigma$. The dual functional has the following property.

\medskip

\begin{lemma}
The dual functional $\mathbb{J}_{\theta}(\Lambda)$ is strictly convex.
\end{lemma}

\begin{proof}
This is equivalent to $\delta^{2} \mathbb{J}_{\theta} > 0$ where
\begin{equation}
\delta^{2} \mathbb{J}_{\theta}(\Lambda; \delta \Lambda)=\int_\mathbb{R} \frac{\theta(x)}{q(x)^{2}}\left(G(x)^{T} \delta\Lambda G(x)\right)^{2}dx
\label{SecondDeriv}
\end{equation}
By (\ref{SecondDeriv}), we have $\delta^{2} \mathbb{J}_{\theta} \geq 0$, so it remains to show that
$$
    \delta^{2} \mathbb{J}_{\theta} > 0, \quad \text{for all $\delta \Lambda \neq \mathbf{0}$},
$$
which follows directly from Lemma \ref{lemma36}, replacing $\Lambda$ by $\delta\Lambda$.
\end{proof}

This leads to the following update of Algorithm 1, which is executed for a particular choice of $\theta$.
\medskip

\begin{algorithm}[htb] 
\caption{Bayesian filtering with density surrogate using power moments at time $t$.} 
\begin{algorithmic}[1] 
\Require 
System parameters: $f_{t}, h_{t}$;
non-Gaussian densities: $\eta_{t}, \epsilon_{t}$;
prediction at time $t-1$: $\rho_{x_{0}}(x) \text{ or } \hat{\rho}_{x_{t} \mid \mathcal{Y}_{t-1}}(x)$;
\Ensure 
Prediction at time $t$: $\hat{\rho}_{x_{t+1} \mid \mathcal{Y}_{t}}(x)$.
\State Calculate $\hat{\rho}_{x_{t}|\mathcal{Y}_{t}}(x)$ by (\ref{Update1}) or (\ref{Update2}); 
\State Calculate $\Sigma$ by (\ref{MomentUpdate});
\State Do the optimization (\ref{LossFunc}) to obtain the order-$2n$ density surrogate of  \eqref{Prediction},
which is the new predictor $\hat{\rho}_{x_{t+1} \mid \mathcal{Y}_{t}}(x)$.
\label{AlgoSurrogate2}
\end{algorithmic}
\label{Algo2}
\end{algorithm}

\section{Tails and error analyses of the proposed filter}
\label{MomentEstimation}

Given a prior probability density $\theta$, Algorithm 2 provides us with a unique solution $\hat{\rho}$ to the truncated Hamburger moment problem, that is, with a unique surrogate probability density $\hat{\rho}$. In this calculation the choice of $n$ may be crucial, as $\hat{\rho}$ may have only a finite number of moments. Indeed, we may want to consider situations when the density has a heavy tail. 
In this section, we establish the conditions on the prior $\theta$ for the density estimate $\hat{\rho}$ to satisfy tail specifications.

\subsection{Light-tailed density surrogate and the moment error propagation}

We first introduce the concept of the sub-Gaussian distributions \cite{vershynin2018high}, which, loosely speaking, are distributions whose tails are dominated by the tails of a Gaussian distribution, i.e., decay at least as fast as a Gaussian.  More precisely, a random variable $X$ is called sub-Gaussian if the moments of $X$ satisfy
\begin{equation}
\|X\|_{L^{p}}=\left(\mathbb{E}|X|^{p}\right)^{1 / p} \leq K_{1} \sqrt{p} \text { for all } p \geq 1
\label{submoment}
\end{equation}
or the moment generating function of $X^{2}$ is bounded at some point, namely
\begin{equation}
\mathbb{E} \left[\exp \left(X^{2} / K_{2}^{2}\right)\right] \leq 2
\label{Pupper}
\end{equation}
where $K_{1}, K_{2} \in \mathbb{R}_{+}$ are two parameters. We denote the space of all sub-Gaussian distributions as $\mathcal{SG}$. Then we have the following theorem.

\medskip

\begin{theorem}
All power moments of the density surrogate $\hat{\rho}$ exist and are finite if and only if the prior $\theta \in \mathcal{SG}$.
\end{theorem}
\begin{proof}
We first prove the necessity. We have
\begin{equation*}
\mathbb{E}\left[ |\hat{x}|^{p} \right]
= \int_{\mathbb{R}}|x|^{p}\hat{\rho}(x)dx
= \int_{\mathbb{R}}|x|^{p}\frac{\theta(x)}{\hat{q}(x)}dx.
\end{equation*}
By Lemma \ref{lemma33}, we have that  $\hat{q}(x) > 0$. We also note that $|x|^{p}, \theta(x)$ are both positive.

Since the prior $\theta \in \mathcal{SG}$, by \eqref{submoment} we have
$$
\begin{aligned}
\int_{\mathbb{R}}|x|^{p}\frac{\theta(x)}{\hat{q}(x)}dx & \leq \frac{1}{\min_{x}\hat{q}(x)}\int_{\mathbb{R}}|x|^{p}\theta(x)dx\\
& \leq \frac{1}{\min_{x}\hat{q}(x)} \left(K_{1} \sqrt{p}\right)^{p} ,
\end{aligned}
$$
which proves that $\mathbb{E}\left[ |\hat{x}|^{p} \right]$, $p= 1,2,3,\dots$, are all finite. However we have $|\mathbb{E}\left[ \hat{x}^{p} \right]| \leq \mathbb{E}\left[ |\hat{x}|^{p} \right]$. Then
$\mathbb{E}\left[ \hat{x}^{p} \right]$ are also finite, and hence all moments exist and are finite.

Next we prove the sufficiency. In view of \eqref{MiniForm},

\begin{equation*}
\begin{aligned}
& \mathbb{E}_{\theta} \left[\exp \left(x^{2} / K_{2}^{2}\right)\right]\\
= & \sum_{i=0}^{\infty} \frac{1}{i!}\int_{\mathbb{R}}\left(x^{2} / K_{2}^{2}\right)^{i}\theta(x)dx\\
= & \sum_{i=0}^{\infty} \frac{1}{i!}\int_{\mathbb{R}}\left(x^{2} / K_{2}^{2}\right)^{i}G(x)^{T} \Lambda G(x)\hat{\rho}(x)dx
\end{aligned}
\end{equation*}
Then, with $\hat{\Lambda}_{j, k}$ being the entries of the matrix $\hat{\Lambda}$, we have
\begin{equation*}
\begin{aligned}
& \mathbb{E}_{\theta} \left[\exp \left(x^{2} / K_{2}^{2}\right)\right]\\
= & \sum_{i=0}^{\infty}\sum_{j=1}^{n}\sum_{k=1}^{n} \frac{\hat{\Lambda}_{j, k}}{i!}\int_{\mathbb{R}}\left(x^{2} / K_{2}^{2}\right)^{i}x^{j+k-2}\hat{\rho}(x)dx\\
= & \sum_{i=0}^{\infty}\sum_{j=1}^{n}\sum_{k=1}^{n} \frac{\hat{\Lambda}_{j, k}}{i!K_{2}^{2i}}\mathbb{E}[\hat{x}^{2i+j+k-2}].
\end{aligned}
\end{equation*}

Since all power moments of $\hat{\rho}$ exist and are finite, it is always possible to choose a $K_{2} \in \mathbb{R}_{+}$ such that
$$
\left|\sum_{j=1}^{n}\sum_{k=1}^{n}\frac{\hat{\Lambda}_{j, k}}{i!K_{2}^{2i}}\mathbb{E}[\hat{x}^{2i+j+k-2}]\right| \leq \frac{1}{2^{i}},\ i \geq 0,
$$
and then we have
$$
\mathbb{E}_{\theta} \left[\exp \left(x^{2} / K_{2}^{2}\right)\right] \leq \frac{1}{1-1/2} = 2,
$$
i.e., the prior $\theta$ is sub-Gaussian by \eqref{Pupper}. This completes the proof of sufficiency.  
\end{proof}

Error propagation through the whole filtering process is a problem in the filter design. Unlike other pdf approximation problems, the estimation is done at each time step $t$, which means that the approximation errors of the previous iterations may have a cumulative effect on the current estimation. 

With the proposed condition on $\theta$, we are always able to ensure the existence and boundedness of all power moments of $\hat{\rho}$. We will first analyze the error propagation of the first $2n$ power moments when $\theta \in \mathcal{SG}$. Since the approximation errors caused by the time updates could effect the measurement updates, we analyze the first $2n$ moment terms of not only $\hat{\rho}_{x_{t+1} \mid \mathcal{Y}_{t}}$ but also $\hat{\rho}_{x_{t} \mid \mathcal{Y}_{t}}$.

\medskip

\begin{theorem}
Suppose $\hat{\rho}_{x_1|\mathcal{Y}_0}$ is a surrogate for $\rho_{x_1|\mathcal{Y}_0}$, and let $\hat{\rho}_{x_t|\mathcal{Y}_t}$ and $\hat{\rho}_{x_{t+1}|\mathcal{Y}_t}$ be obtained from Algorithm 1 for $t = 2, 3, \cdots$. Then the power moments of $\hat{\rho}_{x_t|\mathcal{Y}_t}$ and $\hat{\rho}_{x_{t+1}|\mathcal{Y}_t}$ up to order $2n$ are asymptotically unbiased in $n$ from those of $\rho_{x_t|\mathcal{Y}_t}$ and $\rho_{x_{t+1}|\mathcal{Y}_t}$ respectively and are approximately identical to them for a sufficiently large $n$, given that all power moments of $x_{t}$ and the corresponding $\hat{x}_{t}$ exist and are finite.
\label{theorem42}
\end{theorem}

%
%
%
%
%
%

\begin{proof}
For the sake of simplicity, we omit the normalizing factor in the measurement update equations \eqref{Update1} and \eqref{Update2}, which does not affect the remaining results in this section. The first $2n$ moment terms of $\rho_{x_{1} \mid \mathcal{Y}_{0}}$ are identical to $\hat \rho_{x_{1} \mid \mathcal{Y}_{0}}$ after the first time update, i.e.,
\begin{equation}
    \int_{\mathbb{R}} x^{k} \rho_{x_{1} \mid \mathcal{Y}_{0}} dx = \int_{\mathbb{R}} x^{k} \hat \rho_{x_{1} \mid \mathcal{Y}_{0}} dx, \quad k = 0, \cdots, 2n.
\label{2nEqual}
\end{equation}
Then, referring to \eqref{Update2}, we can write the moment terms of $\rho_{x_{1} \mid \mathcal{Y}_{1}}$ as
\begin{equation*}
    \mathbb{E}\left ( x_{1}^{k}|\mathcal{Y}_{1} \right ) = \int_{\mathbb{R}} x^{k} \rho_{\epsilon_{1}}\left(y_{1}-h_{1} x\right) \rho_{x_{1} \mid \mathcal{Y}_{0}}(x)dx
\end{equation*}
for $k = 0, \cdots, 2n$, and those of $\hat \rho_{x_{1} \mid \mathcal{Y}_{1}}$ as,
\begin{equation*}
    \mathbb{E}\left ( \hat{x}_{1}^{k}|\mathcal{Y}_{1} \right ) = \int_{\mathbb{R}} x^{k} \rho_{\epsilon_{1}}\left(y_{1}-h_{1} x\right) \hat \rho_{x_{1} \mid \mathcal{Y}_{0}}(x)dx
\end{equation*}
for $k = 0, \cdots, 2n$. Therefore we have,
\begin{equation}
\begin{aligned}
    & \mathbb{E}\left ( x_{1}^{k}|\mathcal{Y}_{1} \right ) - \mathbb{E}\left ( \hat{x}_{1}^{k}|\mathcal{Y}_{1} \right )\\
    & = \int_{\mathbb{R}} x^{k} \rho_{\epsilon_{1}}\left(y_{1}-h_{1} x\right) \left ( \rho_{x_{1} \mid \mathcal{Y}_{0}}(x) - \hat \rho_{x_{1} \mid \mathcal{Y}_{0}}(x) \right )dx.
\label{lighttail}
\end{aligned}
\end{equation}

We note that $\rho_{\epsilon_{1}}\left(y_{1}-h_{1} x\right)$ is analytic almost everywhere. Assume $\rho_{\epsilon_{1}}\left(y_{1}-h_{1} x\right)$ is analytic at point $x_{0}$, then it is feasible for us to write the Taylor series at this point. Without loss of generality, we take $x_{0} = 0$, then we have

$$
    \rho_{\epsilon_{1}}\left(y_{1}-h_{1} x\right) = \sum_{i = 0}^{+\infty} \frac{\rho^{(i)}_{\epsilon_{1}}\left(y_{1}\right)}{i!}x^{i}
$$

Since all power moments of $x_{1}$ and $\hat{x}_{1}$ exist and are finite, we have
\begin{equation*}
\begin{aligned}
    & \mathbb{E}\left ( x_{1}^{k}|\mathcal{Y}_{1} \right ) - \mathbb{E}\left ( \hat{x}_{1}^{k}|\mathcal{Y}_{1} \right ) \\
    = & \sum_{i = 0}^{+\infty} \frac{\rho^{(i)}_{\epsilon_{1}}\left(y_{1}\right)}{i!} \int_{\mathbb{R}} x^{k+i} \left ( \rho_{x_{1} \mid \mathcal{Y}_{0}} - \hat \rho_{x_{1} \mid \mathcal{Y}_{0}} \right )dx,
\end{aligned}
\end{equation*}
which, in view of \eqref{2nEqual}, yields
\begin{equation}
\begin{aligned}
    & \mathbb{E}\left ( x_{1}^{k}|\mathcal{Y}_{1} \right ) - \mathbb{E}\left ( \hat{x}_{1}^{k}|\mathcal{Y}_{1} \right ) \\
    = & \sum_{i = 2n-k+1}^{+\infty} \frac{\rho^{(i)}_{\epsilon_{1}}\left(y_{1}\right)}{i!} \int_{\mathbb{R}} x^{k+i} \left ( \rho_{x_{1} \mid \mathcal{Y}_{0}} - \hat \rho_{x_{1} \mid \mathcal{Y}_{0}} \right )dx,\\
   &\qquad\qquad\qquad k=0,1,\dots, 2n,
    \label{momentdiff}
\end{aligned}
\end{equation}
which tends to zero as $n\to \infty$. Thus, by properly selecting a sufficient large $n$, we have
\begin{equation*}
    \mathbb{E}\left ( x_{1}^{k}|\mathcal{Y}_{1} \right ) \approx \mathbb{E}\left ( \hat{x}_{1}^{k}|\mathcal{Y}_{1} \right ),\quad k = 0, \cdots, 2n,
\end{equation*}

Similarly we can prove
\begin{equation*}
    \mathbb{E}\left ( x_{t}^{k}|\mathcal{Y}_{t} \right ) \approx \mathbb{E}\left ( \hat{x}_{t}^{k}|\mathcal{Y}_{t} \right ),\quad k = 0, \cdots, 2n,
\end{equation*}
and
\begin{equation*}
    \mathbb{E}\left ( x_{t+1}^{k}|\mathcal{Y}_{t} \right ) \approx \mathbb{E}\left ( \hat{x}_{t+1}^{k}|\mathcal{Y}_{t} \right ),\quad k = 0, \cdots, 2n,
\end{equation*}
as claimed.
\end{proof}

Theorem \ref{theorem42} proves that the first $2n$ moment terms of the estimated densities with the density surrogate are approximately the true ones throughout the whole filtering process for sufficiently large $n$, i.e., $\hat \rho_{x_{t+1} \mid \mathcal{Y}_{t}}$, and $\hat \rho_{x_{t} \mid \mathcal{Y}_{t}}$ are approximately order-$2n$ density surrogate of $\rho_{x_{t+1} \mid \mathcal{Y}_{t}}$ and $\rho_{x_{t} \mid \mathcal{Y}_{t}}$. It reveals the fact that approximation using the truncated power moments does not introduce uncontrollable cumulative errors to the first $2n$ moment terms of the estimated pdfs, given that all power moments of the true system states exist and are finite, i.e., the prior $\theta \in \mathcal{SG}$.

\subsection{Heavy-tailed density surrogate and the moment errors}
We have proposed the feasible class of $\theta$ to ensure the existence and boundedness of the power moments of the density surrogate $\hat{\rho}(x)$. However in some situations, one desires state estimates with heavy tails. In the following part of this section, we analyze the error propagation of the power moments, given that the power moments are not all finite, i.e., $\theta \notin \mathcal{SG}$.

The calculation \eqref{momentdiff} is still valid, but, since the power moments of $\hat{\rho}$ are not all finite, i.e.,
$$
\int_{\mathbb{R}} x^{k} \left ( \rho_{x_{t+1} \mid \mathcal{Y}_{t}} - \hat \rho_{x_{t+1} \mid \mathcal{Y}_{t}} \right )dx
$$
may be infinite for some $k$, we cannot draw the same conclusion as in the light-tailed case. However, we note that
\begin{equation*}
\begin{aligned}
    & \left|\mathbb{E}\left ( x_{t+1}^{k}|\mathcal{Y}_{t+1} \right ) - \mathbb{E}\left ( \hat{x}_{t+1}^{k}|\mathcal{Y}_{t+1} \right )\right|\\
    = & \left|\int_{\mathbb{R}} x^{k} \rho_{\epsilon_{t+1}}\left(y_{t+1}-h_{t+1} x\right) \left ( \rho_{x_{t+1} \mid \mathcal{Y}_{t}} - \hat \rho_{x_{t+1} \mid \mathcal{Y}_{t}} \right )dx\right|\\
    \leq & \int_{\mathbb{R}} \left| x \right|^{k}\rho_{\epsilon_{t+1}}\left(y_{t+1}-h_{t+1} x\right)\left|  \rho_{x_{t+1} \mid \mathcal{Y}_{t}} - \hat \rho_{x_{t+1} \mid \mathcal{Y}_{t}} \right|dx,
\end{aligned}
\label{Heavytail}
\end{equation*}
and therefore
$$
\begin{aligned}
& \left|\mathbb{E}\left ( x_{t+1}^{k}|\mathcal{Y}_{t+1} \right ) - \mathbb{E}\left ( \hat{x}_{t+1}^{k}|\mathcal{Y}_{t+1} \right )\right|\\
&\quad \qquad\leq  C_{k}\max_{x}\left|  \rho_{x_{t+1} \mid \mathcal{Y}_{t}} - \hat \rho_{x_{t+1} \mid \mathcal{Y}_{t}} \right|,
\end{aligned}
$$
where $C_{k} := \int_{\mathbb{R}} \left| x \right|^{k}\rho_{\epsilon_{t+1}}\left(y_{t+1}-h_{t+1} x\right)dx$ is a constant unrelated to $\hat \rho_{x_{t+1} \mid \mathcal{Y}_{t}}$.
Consequently, we have proven the following theorem.

\medskip

\begin{theorem}
The errors of the power moments of $\hat{\rho}_{x_{t+1}|\mathcal{Y}_{t+1}}$ are each bounded by a value which is proportional to the $L_{\infty}$ norm of the error of the density surrogate $\hat \rho_{x_{t+1} \mid \mathcal{Y}_{t}}$. 
\label{theorem43}
\end{theorem}

Theorem \ref{theorem43} reveals the fact that with a satisfactory performance of density estimation, i.e., a relatively small $\max_{x}\left|  \rho_{x_{t+1} \mid \mathcal{Y}_{t}} - \hat \rho_{x_{t+1} \mid \mathcal{Y}_{t}} \right|$, the error of the estimated moments of the density is also small. 

We have analyzed the error propagation of estimated power moments with light and heavy tailed density surrogates. But we note that in real applications, it is not possible for us to treat the infinite-dimensional estimation problem, i.e., to use the full power moment sequence for density estimation. Then the density estimate is not always identical to the true density for either $\theta \in \mathcal{SG}$ or $\theta \notin \mathcal{SG}$. In the next section, we propose to analyze the error upper bounds of $\hat{\rho}$ to reveal its maximum difference from the true one, given the first $2n$ terms of power moments.

\subsection{Error upper bounds of the density surrogate}
\label{QuantAnalysis}

To our knowledge, an error upper bound for the state estimate has not been established in Bayesian filtering with non-Gaussian distributions. The reason is that a continuous form of parametrization of the system state has not been proposed. In this section, we propose an error upper bound of $\hat{\rho}(x)$ in the sense of total variation distance, which is a measure widely used in the moment problem \cite{Aldo2003A, tagliani2003maximum}. This upper bound distinguishes our proposed filter from other Bayesian filters.

The total variation distance between the density estimate $\hat{\rho}$ and the true density $\rho$ is defined as follows:
\begin{equation}
\begin{aligned}
    V(\hat\rho, \rho) & = \sup_{x} \left|\int_{\left(-\infty, x \right] }(\hat \rho - \rho) d x\right|\\
    & = \sup_{x} \left| F_{\hat \rho} - F_{\rho} \right| 
\end{aligned}
\label{lim}    
\end{equation}
where $F_{\hat \rho}$ and $F_{\rho}$ are the two cumulative distribution functions of $\hat \rho$ and $\rho$.

In \cite{Aldo2003A}, Shannon-entropy is used to calculate the upper bound of the total variation distance. The Shannon-entropy \cite{shannon1948mathematical} is defined as
$$
    H[\rho] = - \int_{\mathbb{R}}\rho(x) \log \rho(x)dx.
$$
We first introduce the Shannon-entropy maximizing distribution $F_{\breve{\rho}}$, of which the moments are calculated by \eqref{MomentUpdate}. It has the following density function \cite{kapur1992entropy},
\begin{equation}
\label{rhobreve}
    \breve{\rho}(x) = \exp \left ( - \sum_{i = 0}^{2n} \lambda_{i} x^{i} \right )
\end{equation}where $\lambda_{0}, \cdots, \lambda_{2n}$ are determined so that $ \breve{\rho}$ has the same moments $\hat{\sigma}_1,\hat{\sigma}_2, \dots,\hat{\sigma}_{2n}$ as the density $\hat{\rho}$, i.e.,
$$
    \int_{\mathbb{R}} x^j  \breve{\rho}(x) dx=\hat{\sigma}_{j}, \quad j=0,1, \cdots, 2n .
$$
Hence
$$
H[\breve\rho]=\sum_{i = 0}^{2n} \lambda_{i}\int_\mathbb{R} x^{i}\breve{\rho}(x)dx  =\sum_{i = 0}^{2n} \lambda_{i}\hat{\sigma}_i
$$
Then, following \cite{Aldo2003A}, we form the KL distance between the true density and the Shannon-entropy maximizing density, i.e., in view of \eqref{rhobreve},
$$
\begin{aligned}
    \mathbb{KL} \left(\rho \| \breve{\rho}\right) & =  \int_{\mathcal{\mathbb{R}}} \rho(x) \log \frac{\rho(x)}{\breve{\rho}(x)} dx \\
    & =- H\left [ \rho \right ] + \sum_{i = 0}^{2n} \lambda_{i}\sigma_i,
\end{aligned}
$$
However, if $\theta \in \mathcal{SG}$ and $n$ is sufficiently large, $\hat{\sigma}_i$ is approximately equal to $\sigma_i$ for $i=0,1,\dots2n$
($\sigma_i\approx \hat{\sigma}_{i}$) by Theorem \ref{theorem42}, and hence
$$
 \mathbb{KL} \left(\rho \| \breve{\rho}\right) \approx  H [ \breve{\rho}] - H[\rho].
$$
Similarly, we obtain
$$
    \mathbb{KL} \left(\hat{\rho} \| \breve{\rho}\right)\approx H [\breve{\rho}] - H[\hat{\rho}].
$$
 By \cite{1970Correction, Aldo2003A}, we have
$$
\begin{aligned}
    V \left ( \breve{\rho}, \hat{\rho}\right ) & \leq 3\left[-1+\left\{1+\frac{4}{9} \mathbb{KL} \left(\hat{\rho} \| \breve{\rho} \right)\right\}^{1 / 2}\right]^{1 / 2} \\
    & = 3\left[-1+\left\{1+\frac{4}{9} \left ( H\left [ \breve{\rho} \right ] - H\left [ \hat{\rho} \right ] \right )\right\}^{1 / 2}\right]^{1 / 2}
\label{Vbound}
\end{aligned}
$$
and
$$
    V \left ( \breve{\rho}, \rho \right ) \leq 3\left[-1+\left\{1+\frac{4}{9} \left ( H\left [ \breve{\rho} \right ] - H\left [ \rho \right ] \right )\right\}^{1 / 2}\right]^{1 / 2}
$$
Then we obtain the upper bound of the error
$$
\begin{aligned}
& V \left ( \hat{\rho}, \rho \right ) \\
= & \sup_{x}|F_{\hat{\rho}}\left ( x \right )-F_{\rho}\left ( x \right )| \\
\leq & \sup_{x}\left(\left|F_{\hat{\rho}}\left ( x \right )-F_{\breve{\rho}}\left ( x \right )\right|+\left|F_{\breve{\rho}}(x)-F_{\rho(x)}\right|\right) \\
\leq & \sup_{x} \left|F_{\hat{\rho}}\left ( x \right )-F_{\breve{\rho}}\left ( x \right )\right|+\sup_{x} \left|F_{\breve{\rho}}(x)-F_{\rho(x)}\right| \\
\leq & 3 \left[-1+\left\{1+\frac{4}{9}\left(H\left [ \breve{\rho} \right ] - H\left [ \hat{\rho} \right ]\right)\right\}^{1 / 2}\right]^{1 / 2} \\
+ & 3\left[-1+\left\{1+\frac{4}{9}\left(H\left [ \breve{\rho} \right ] - H\left [ \rho \right ]\right)\right\}^{1 / 2}\right]^{1 / 2}
\end{aligned}
\label{UpperBoundUnbiased}
$$

In some practical situations, e.g. financial applications, error upper bounds of the probability of the state estimate within intervals, e.g. $\left|P(x_{t}\geq a) - P(\hat{x}_{t}\geq a)\right|$, $\left|P(a \leq x_{t} \leq b) - P(a \leq \hat{x}_{t} \leq b)\right|$, are desired for people to make conservative decisions. However to our knowledge, there has not been a non-Gaussian Bayesian filter which provides such kinds of tight bounds without assuming the density functions to fall within specific classes.

In Section \ref{MomentEstimation}, we have proved that the power moments of the density estimates are approximately the true ones, by using our proposed algorithm when the density surrogate is light-tailed. The estimation error of the moments have also been proved to be bounded and small with satisfactory density estimation performance, which can be achieved by our proposed algorithm. We note that there are a series of research results on the tight bounds of the moment problem. These results make it feasible for us to derive upper bounds for the density estimates during the filtering process by our proposed algorithm. For example, achievable upper bounds $\max P(x_{t}\geq a)$ and $\max P(a \leq x_{t} \leq b)$ given the moment constraints are proposed in \cite{bertsimas2005optimal}. By these upper bounds, we can then obtain the upper bounds of errors
$$
\begin{aligned}
& \left|P(x_{t}\geq a) - P(\hat{x}_{t} \geq a)\right|\\
\leq & \max \left\{\max P(x_{t}\geq a) - P(\hat{x}_{t}\geq a), P(\hat{x}_{t}\geq a) \right\},   
\end{aligned}
$$
and
$$
\begin{aligned}
& \left|P(a \leq x_{t} \leq b) - P(a \leq \hat{x}_{t} \leq b)\right|\\
\leq & \max \left\{ \max P(a \leq x_{t} \leq b) - P(a \leq \hat{x}_{t} \leq b), \right. \\
& \left. P(a \leq \hat{x}_{t} \leq b) \right\}.
\end{aligned}
$$
In conclusion, we have performed quantitative error analyses of the state estimates. An error upper bound of the state estimate in the sense of total variation distance, together with two error upper bounds for the probability of subsets of the real line given the power moments have been proposed in this section. 

\section{Simulation details and results}
\label{SimulationDetails}

In the previous sections, a Bayesian filter with the density parameterized by using the power moments has been proposed. However, there are still several details to note when implementing the filter. This will be done in this section, where we will provide simulation results to validate the filter we propose.

The first problem is the choice of the prior $\theta(x)$. For light-tailed density surrogates, $\theta(x) \in \mathcal{SG}$ can usually be chosen as a Gaussian density function. It ensures that the first $2n$ power moments of $\hat{\rho}(x)$ are finite. Therefore, the problem reduces to determining the mean and variance of the Gaussian distribution. 

At each time step, the first and second order power moments, i.e.,  $\sigma_{1}, \sigma_{2}$ of the density to be estimated can be calculated by (\ref{MomentUpdate}). In practice, we can choose $m = \sigma_{1}$ and $\sigma^{2} > \sigma_{2}$ and determine the prior density $\theta(x) = \mathcal{N}(m, \sigma^{2})$ for each time step. Here we note that a relatively large variance $\sigma^{2}$ is to better adjust to the densities with multiple modes.

Second, we considers the choice of $\rho_{x_{0}}$. In some scenarios, the true probability density of the initial state $x_{0}$ is known prior. For scenarios where the initial state $x_{0}$ is not known prior, we take an arbitrary moment sequence $\bar{\sigma}_{2n}$ which satisfies that the Hankel matrix is positive definite. Then the $\rho_{x_{0}}$ can be obtained by doing the optimization (\ref{LossFunc}) given the moment constraints $\bar{\sigma}_{2n}$.

We note that the density $\rho_{x_{t+1} \mid \mathcal{Y}_{t}}(x)$ does not always have an explicit function form, i.e., it is not always possible to obtain the true system states. It makes comparing the estimates of the density to the true ones infeasible. However, we note that when $\eta_{t}$ is a discrete random variable,  the density $\rho_{x_{t+1} \mid \mathcal{Y}_{t}}(x)$ can be written as

\begin{equation}
    \rho_{x_{t+1} \mid \mathcal{Y}_{t}}(x) = \sum_{i=1}^{m} \rho _{i} \cdot \rho_{x_{t}\mid \mathcal{Y}_{t}}\left(\frac{x-\xi_{i}}{f_{t}}\right)
\end{equation}
which is a mixture of densities and is Lebesgue integrable (analytic if $\rho_{x_{t+1} \mid \mathcal{Y}_{t}}$ is analytic). In order to compare the density estimates to the true density for validating the performance of the proposed surrogates, we simulate the mixture of densities in the following parts of this section. Moreover, the average estimation error of $\hat{\rho}$ is calculated by the total variation distance $V \left( \hat{\rho}, \rho \right)$, i.e., \eqref{lim} in the following simulations.

\subsection{Density estimation with different number of moment terms}

In this part of section, we simulate on density estimation with different number of moment terms. In Example 1, we choose the true density to be a mixture of two Gaussians where there are two modes

$$
     \rho(x) = \frac{0.3}{\sqrt{2\pi}}e^{\frac{(x-2)^{2}}{2}} + \frac{0.7}{\sqrt{2\pi}}e^{\frac{(x+2)^{2}}{2}}.
$$

\begin{figure}[htbp]
\centering
\includegraphics[scale=0.38]{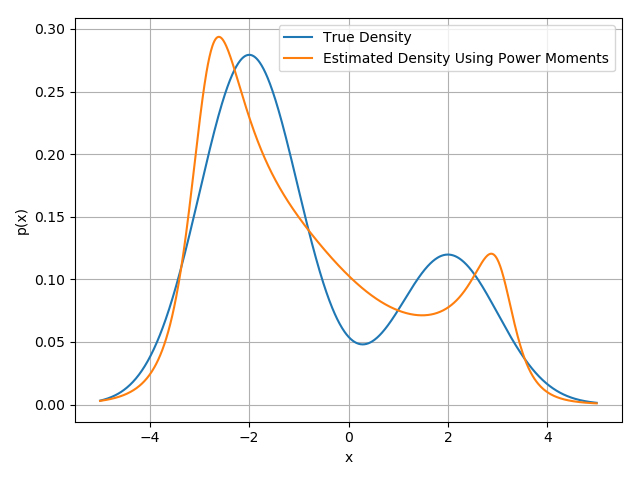}
\centering
\caption{Simulation result of Example 1. The blue curve represents the true density function. The orange one represents the density estimate using power moments.}
\label{fig1}
\end{figure}

The function class of the true density is not known prior. We use power moments up to order $6$ to estimate the density function. $\theta(x)$ is chosen as $\mathcal{N}(-0.8, 3^{2})$. The highest order of the polynomial $q(x)$ is $6$. The density estimate $\hat{\rho}(x) = \theta(x) / q(x)$, where $q(x) = 2.30\cdot 10^{-3}x^{6} + 3.02\cdot 10 ^{-3}x^{5} - 2.55 \cdot 10^{-2}x^{4} - 6.58 \cdot 10^{-2}x^{3} - 4.10 \cdot 10^{-2}x^{2} + 3.58 \cdot 10^{-1}x + 1.25$. The estimated density function is given in Figure \ref{fig1}. We note that the two modes are well estimated by the parametrization. The estimation error is $V(\hat{\rho}, \rho) = 0.0331$. 

\begin{figure}[htbp]
\centering
\includegraphics[scale=0.38]{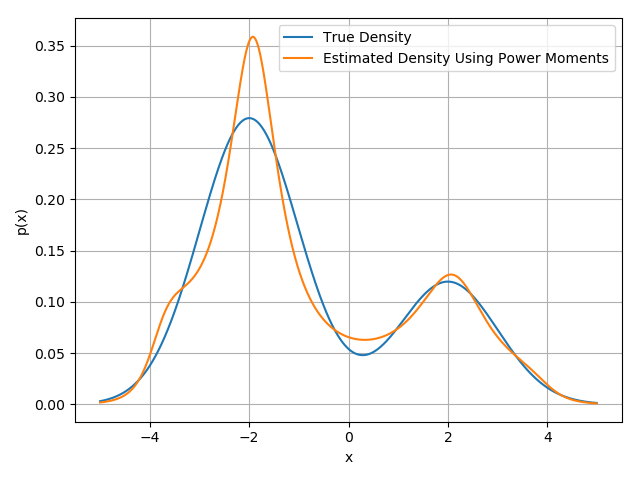}
\centering
\caption{Simulation result of Example 2.}
\label{fig2}
\end{figure}

In Example 2, we simulate the same true density as in Example 1. However, the highest order of moments used is $8$ in this example. $\theta(x)$ is chosen as $\mathcal{N}(-0.8, 3^{2})$. The density estimate $\hat{\rho}(x) = \theta(x) / q(x)$, where $q(x) = 3.81\cdot 10^{-4}x^{8} - 2.46\cdot 10 ^{-4}x^{7} - 1.37\cdot 10^{-2}x^{6} + 8.43\cdot 10 ^{-3}x^{5} + 1.74 \cdot 10^{-1}x^{4} - 8.88 \cdot 10^{-2}x^{3} - 8.64 \cdot 10^{-1}x^{2} + 3.20 \cdot 10^{-1}x + 1.96$. The estimated density function is given in Figure \ref{fig2}. The estimation error is $V(\hat{\rho}, \rho) = 0.0208$. We note that by using power moments up to order $8$, the result is better than that of order $6$ in the sense of the total variation distance. 

In conclusion, these two simulation results give an example that with higher order moments used, the error of density estimation is less significant, which validates the approximation in Theorem \ref{theorem42}. In the following part of section, we will give simulation results on different types of density functions by our proposed algorithm.

\subsection{Estimation of mixtures of different types of density functions}

In the first two examples, we simulate a mixture of a Gaussian and a Laplacian, and the mixtures of Laplacians.

\begin{figure}[htbp]
\centering
\includegraphics[scale=0.38]{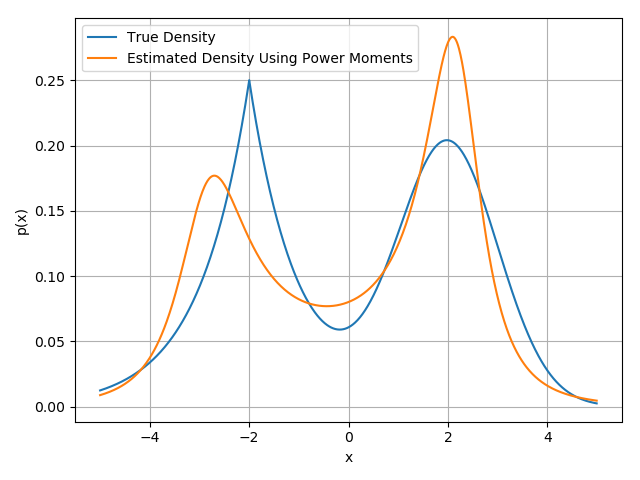}
\centering
\caption{Simulation result of Example 3. The blue curve represents the true density function. The orange one represents  the  density  estimate  using  only  power  moments.}
\label{fig3}
\end{figure}

Example 3 is a bimodal density which is a mixture of a Gaussian and a Laplacian. The probability density function is
$$
     \rho(x) = \frac{0.5}{\sqrt{2\pi}}e^{\frac{(x-2)^{2}}{2}} + \frac{0.5}{2}e^{-\left| x + 2\right|}.
$$
$\theta(x)$ is chosen as $\mathcal{N}(0, 5^{2})$. The highest order of the polynomial $q(x)$ is $4$. We obtain the density estimate $\hat{\rho}(x) = \theta(x) / q(x)$, where $q(x) = 0.0203x^{4} + 0.0280x^{3} - 0.2252x^{2} - 0.1892x + 0.9948$. The simulation result is given in Figure \ref{fig3}. In this example, we note that even if the modes are of different types of distributions, the proposed density surrogate can still treat the density approximation without prior knowledge of the type of distributions. The two distinct modes are well estimated. $V(\hat{\rho}, \rho) = 0.0567$, which is a promising result.

Example 4 is chosen as a bimodal density which is a mixture of two Laplacians. The probability density function is
\begin{equation*}
     \rho(x) = \frac{0.7}{2}e^{-\left| x - 1\right|} + \frac{0.3}{2}e^{-\left| x + 3\right|}.    
\end{equation*}

\begin{figure}[htbp]
\centering
\includegraphics[scale=0.38]{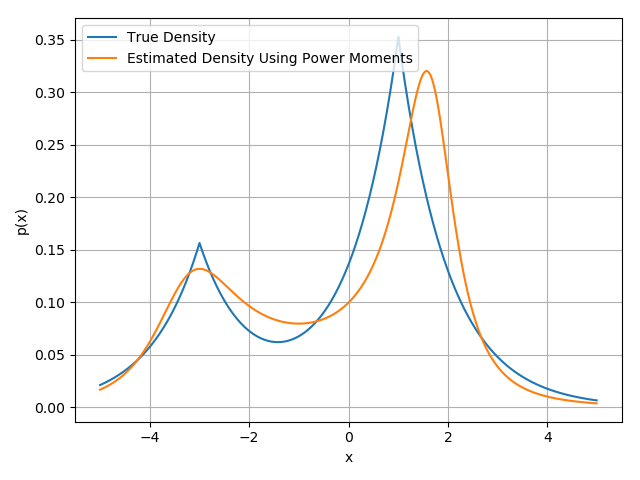}
\centering
\caption{Simulation result of Example 4.}
\label{fig4}
\end{figure}

$\theta(x)$ is chosen as $\mathcal{N}(-0.2, 7^{2})$. The highest order of the polynomial $q(x)$ is $4$. The density estimate $\hat{\rho}(x) = \theta(x) / q(x)$, where $q(x) = 0.0147x^{4} + 0.0476x^{3} - 0.0995x^{2} - 0.2721x + 0.5713$. The simulation result is given in Figure \ref{fig4}. We note that the two modes are well characterized by the density surrogate, even when one has a relatively small probability. The estimation error is $V(\hat{\rho}, \rho) = 0.0744$.

The densities in the two examples above have two modes (peaks). In the following examples, we simulate densities with more modes to validate the performance of our proposed density surrogate.

\begin{figure}[htbp]
\centering
\includegraphics[scale=0.38]{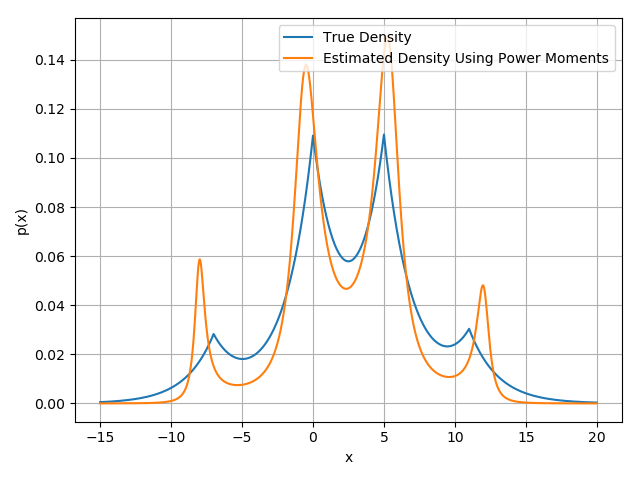}
\centering
\caption{Simulation result of Example 5.}
\label{fig5}
\end{figure}

Example 5 is chosen as a density with four modes which is a mixture of four Laplacians. The probability density function is
$$
\begin{aligned}
     \rho(x) & = \frac{0.4}{2}e^{\left| x \right|} + \frac{0.4}{2}e^{-\left| x - 5\right|} + \frac{0.1}{2}e^{-\left| x + 7\right|}\\
     & + \frac{0.1}{2}e^{-\left| x - 11\right|}.    
\end{aligned}
$$
$\theta(x)$ is chosen as $\mathcal{N}(0.5, 20^{2})$. The highest order of the polynomial $q(x)$ is $8$. The density estimate is $\hat{\rho}(x) = \theta(x) / q(x)$, where $q(x) = 4.22 \cdot 10^{-7}x^{8} - 7.41\cdot 10^{-6}x^{7} - 3.48\cdot 10^{-5}x^{6} + 9.86\cdot 10 ^{-4}x^{5} - 5.48 \cdot 10^{-4}x^{4} - 3.35 \cdot 10^{-2}x^{3} + 7.55 \cdot 10^{-2}x^{2} + 9.69 \cdot 10^{-2}x + 1.70\cdot 10^{-1}$. The simulation result is given in Figure \ref{fig5}.

The simulation result of Example 5 shows the performance of our proposed density surrogate in estimating the multi-modal density without prior knowledge on the modes or function type. The number of modes are correctly observed and the estimation error is satisfactory. The estimation error is $V(\hat{\rho}, \rho) = 0.053$.

\begin{figure}[htbp]
\centering
\includegraphics[scale=0.38]{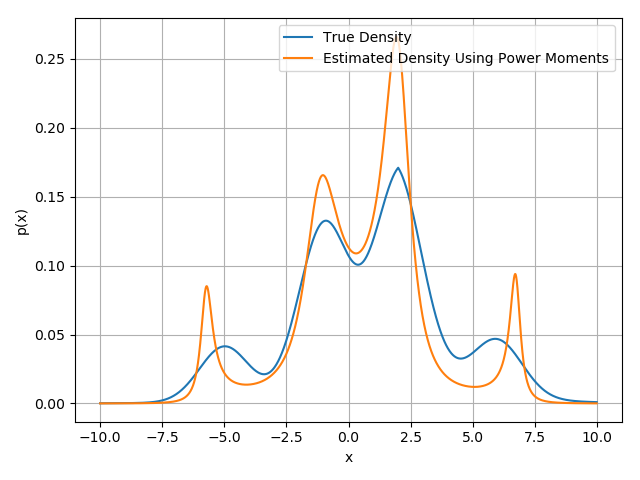}
\centering
\caption{Simulation result of Example 6.}
\label{fig6}
\end{figure}

Example 6 is chosen as a density with four modes which is a mixture of four Gaussians and a Laplacian. The probability density function is
$$
\begin{aligned}
     \rho(x) & = \frac{0.3}{\sqrt{2\pi}}e^{\frac{(x - 2)^{2}}{2}} + \frac{0.3}{\sqrt{2\pi}}e^{\frac{(x + 1)^{2}}{2}} + \frac{0.1}{\sqrt{2\pi}}e^{\frac{(x - 6)^{2}}{2}}\\
     & + \frac{0.1}{\sqrt{2\pi}}e^{\frac{(x + 5)^{2}}{2}} + \frac{0.2}{2}e^{-\left| x - 2\right|}.    
\end{aligned}
$$
It is a complicated mixture of densities. $\theta(x)$ is chosen as $\mathcal{N}(0.6, 10^{2})$. The highest order of the polynomial in the denominator is $8$. The density estimate is $\hat{\rho}(x) = \theta(x) / q(x)$, where $q(x) = 2.31 \cdot 10^{-5}x^{8} - 8.35\cdot 10^{-5}x^{7} - 1.76\cdot 10^{-3}x^{6} + 4.83\cdot 10 ^{-3}x^{5} + 3.75 \cdot 10^{-2}x^{4} - 6.64 \cdot 10^{-2}x^{3} - 1.22 \cdot 10^{-1}x^{2} + 8.95 \cdot 10^{-2}x + 3.52\cdot 10^{-1}$. The simulation result is given in Figure \ref{fig6} and $V \left( \hat{\rho}, \rho \right) = 0.096$. We note that the four modes of the state are well observed and the performance is satisfactory without prior knowledge of the density $\rho_{x_{t+1}|\mathcal{Y}_{t}}(x)$. This example validates the ability of the proposed filter in estimating the complicated densities of the state.

Filtering problems where the state and noise distributions are heavy-tailed is a recent interest of the control community \cite{hanzon2001state, zhu2021novel}. By using the Bayesian filter we propose, it is feasible to treat this problem by choosing $\theta(x)$ as a heavy-tailed distribution. In the following example, we simulate mixtures of student-t distributions. 

Example 7 is chosen as a mixture of two student-t distributions. The probability density function is

$$
\begin{aligned}
     \rho(x) & = \frac{0.4 \cdot 3}{8\left(1+\frac{(x - 2)^{2}}{4}\right)^{\frac{5}{2}}} + \frac{0.6 \cdot 8}{3 \pi \sqrt{5}\left(1+\frac{(x + 2)^{2}}{5}\right)^{3}}  
\end{aligned}
$$

\begin{figure}[htbp]
\centering
\includegraphics[scale=0.38]{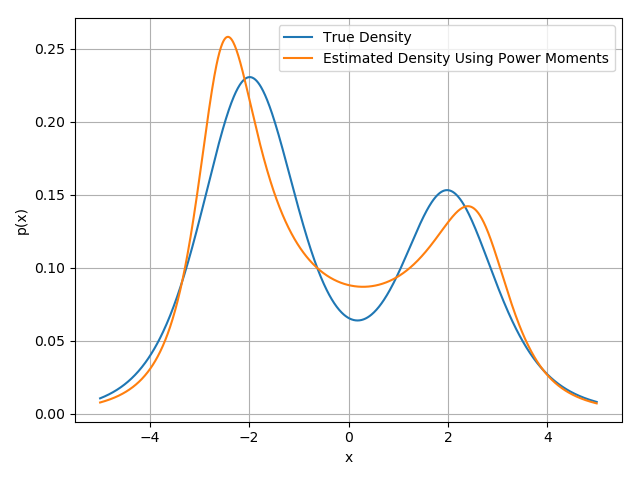}
\centering
\caption{Simulation result of Example 7.}
\label{fig7}
\end{figure}

$\theta(x)$ is chosen as $\mathcal{C}(-0.4, 5)$, where $\mathcal{C}$ denotes the Cauchy distribution. The highest order of the polynomial $q(x)$ is $4$. The density estimate is $\hat{\rho}(x) = \theta(x) / q(x)$, where $q(x) = 0.0114x^{4} - 0.0028x^{3} - 0.1424x^{2} + 0.043x + 0.7180$. The simulation result is given in Figure \ref{fig7}. The estimation error is $V(\hat{\rho}, \rho) = 0.032$. Example 7 validates the ability of our proposed Bayesian filter to treat the heavy-tailed filtering problem.

\section{Conclusion}

In this paper, we propose the use of power moments to parameterize the state of a Bayesian filter considering the first order system. The proposed parametrization is able to characterize a much wider class of density functions without prior knowledge of the density of the state $x_{t}$, e.g. the number of modes and the feasible function class. It is not required to store massive estimates of the state at discrete points. We formulate the density problem as a formal Hamburger moment problem. The existence of solutions to the moment problem is shown, and a Hankel matrix representation of it is proposed. The solutions of the proposed parametrization can be obtained by a convex optimization scheme. The solution to the optimization problem is proved to exist and be unique by proving that the map from the parameters to the power moments is homeomorphic. We prove that all moments of the density surrogate $\hat{\rho}$  exist and are finite if and only if $\theta$ is sub-Gaussian. Given that $\theta$ is sub-Gaussian, we prove that the estimated power moments are asymptotically unbiased and approximately the true ones throughout the filtering process. Therefore by selecting a large enough $n$, there are not severe cumulative errors in our proposed Bayesian filter. We also provide the upper bound of $\hat{\rho}$ when $\theta \notin \mathcal{SG}$. Error upper bounds of the state estimate are also proposed. In the simulation, we simulate mixtures of different types of density functions, including 
Gaussian, Laplacian and student-t. The simulation results on the mixture of student-t distributions validates the ability of the proposed algorithm to treat the heavy-tailed filtering problem, which is a current key problem of stochastic filtering. In future work, we plan to extend our results to the multidimensional systems. The extension is non-trivial, since the parametrization of a multivariate density function given the moment constraints is an open problem. Existence and uniqueness of solution, together with a Positivstellens\"atze of the parametrization, need to be proposed.
\bibliographystyle{plain}
\bibliography{autosam}

\newpage

\begin{wrapfigure}{l}{20mm} 
\includegraphics[width=1in,height=1.25in,clip,keepaspectratio]{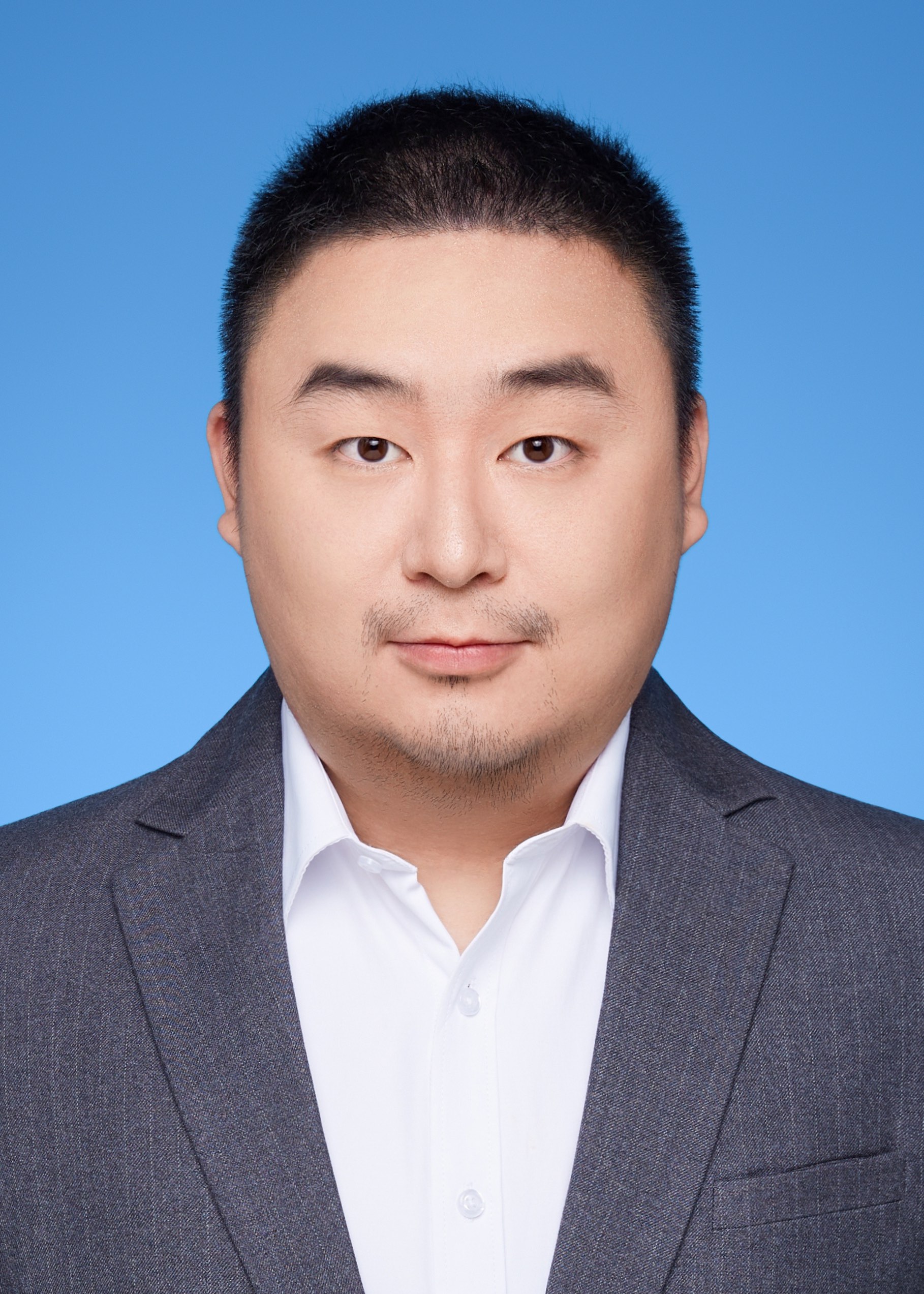}
  \end{wrapfigure}\par
  \textbf{Guangyu Wu} received the B.E. degree from Northwestern Polytechnical University, Xi’an, China, in 2013, and two M.S. degrees, one in control science and engineering from Shanghai Jiao Tong University, Shanghai, China, in 2016, and the other in electrical engineering from the University of Notre Dame, South Bend, USA, in 2018. 

He is currently pursuing the Ph.D. degree at Shanghai Jiao Tong University. His research interests are the moment problem and its applications to stochastic filtering, density steering and statistics. 

\begin{wrapfigure}{l}{25mm} 
\includegraphics[width=1in,height=1.25in,clip,keepaspectratio]{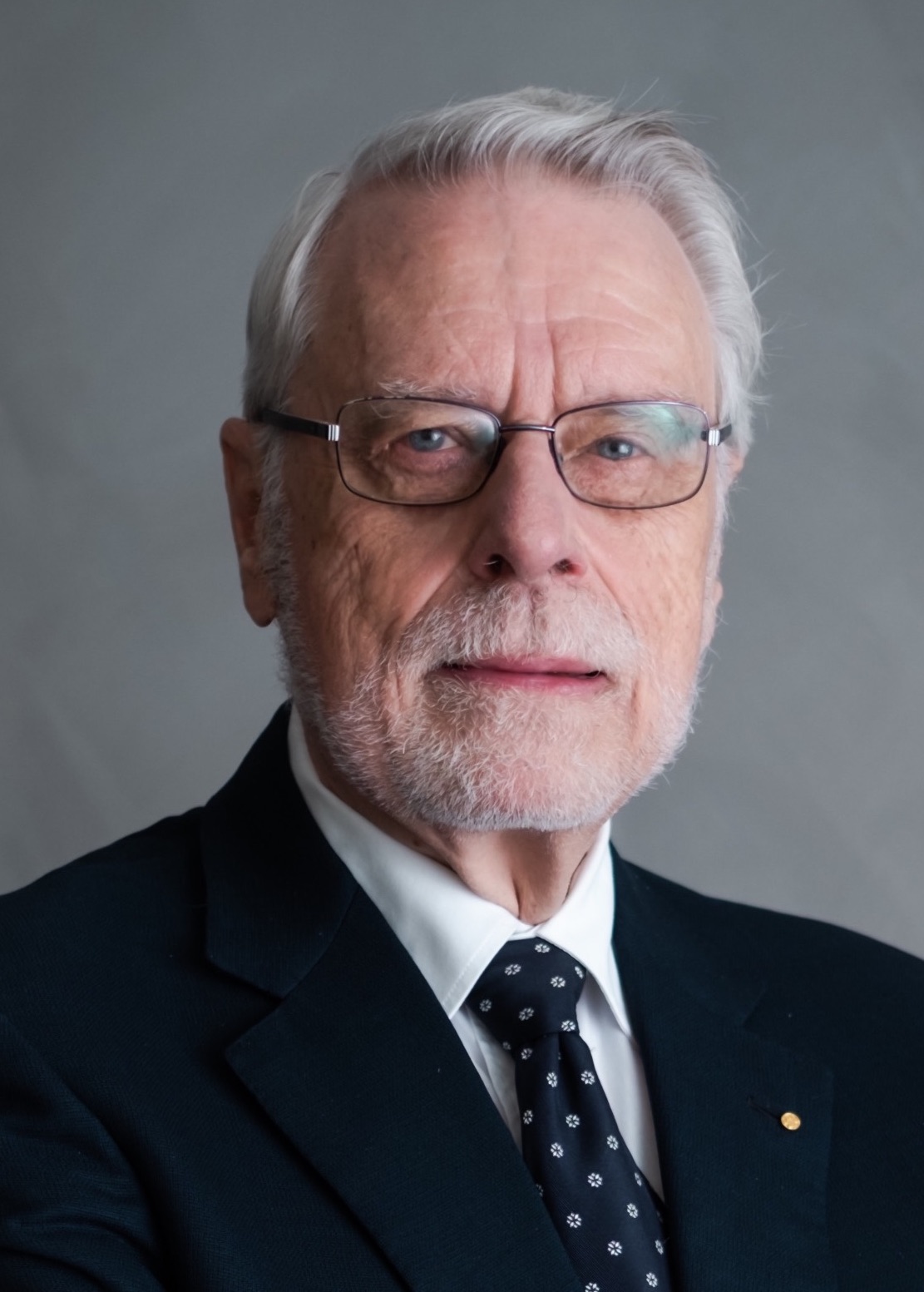}
  \end{wrapfigure}\par
  \textbf{Anders Lindquist} received the Ph.D. degree in optimization and systems theory from the Royal Institute of Technology (KTH), Stockholm, Sweden, in 1972, an honorary doctorate (Doctor Scientiarum Honoris Causa) from Technion (Israel Institute of Technology) in 2010 and Doctor Jubilaris from KTH in 2022.

He is currently a Zhiyuan Chair Professor at Shanghai Jiao Tong University, China, and Professor Emeritus at the Royal Institute of Technology (KTH), Stockholm, Sweden. Before that he had a full academic career in the United States, after which he was appointed to the Chair of Optimization and Systems at KTH.
Dr. Lindquist is a Member of the Royal Swedish Academy of Engineering Sciences, a Foreign Member of the Chinese Academy of Sciences, a Foreign Member of the Russian Academy of Natural Sciences, a Member of Academia Europaea (Academy of Europe), an Honorary Member the Hungarian Operations Research Society, a Fellow of SIAM, and a Fellow of IFAC. He received the 2003 George S. Axelby Outstanding Paper Award, the 2009 Reid Prize in Mathematics from SIAM, and the 2020 IEEE Control Systems Award, the IEEE field award in Systems and Control.

\end{document}